# AR-PPF: Advanced Resolution-Based Pixel Preemption Data Filtering for Efficient Time-Series Data Analysis

Taewoong Kim[1], Kukjin Choi[1], and Sungjun Kim[1]

(Abstract) With the advent of automation, many manufacturing industries have transitioned to data-centric methodologies, giving rise to an unprecedented influx of data during the manufacturing process. This data has become instrumental in analyzing the quality of manufacturing process and equipment. Engineers and data analysts, in particular, require extensive time-series data for seasonal cycle analysis. However, due to computational resource constraints, they are often limited to querying short-term data multiple times or resorting to the use of summarized data in which key patterns may be overlooked. This study proposes a novel solution to overcome these limitations; the advanced resolution-based pixel preemption data filtering (AR-PPF) algorithm. This technology allows for efficient visualization of time-series charts over long periods while significantly reducing the time required to retrieve data. We also demonstrates how this approach not only enhances the efficiency of data analysis but also ensures that key feature is not lost, thereby providing a more accurate and comprehensive understanding of the data.

## 1 Introduction

In recent years, many manufacturing companies have been collecting massive amounts of sensor data to improve productivity as they configure smart factories [1]. The collected data is loaded into systems such as Equipment Engineering System (EES) and engineers visualize the sensor data to make decisions [2].

In short-term periods, time-series data can be efficiently visualized by querying the entire dataset. However, as the query period extends, the rendering time for the final chart increases proportionally. This increase occurs because, although the query time remains constant, the total amount of sensor data grows with the extended sampling period, thereby increasing the rendering time (Fig. 1). To mitigate this delay, it is effective to eliminate visually redundant data points, thus reducing the total data volume. This approach allows for faster rendering while maintaining a visually comparable chart.

A few years ago, we developed an algorithm named "Resolution-based Pixel Preemption Filtering (R-PPF)" for data filtering, which can be utilized in various charting methods within analytic systems. This algorithm selectively retains data points based on the bucket size set on the screen, preemptively omitting all others. This process results in minimal visual difference while significantly reducing the data volume for rendering. Fig. 2 illustrates the data filtering process using the R-PPF algorithm. It assesses the priority of pixels to determine areas of interest and regions that require filtering. Based on the pre-defined a bucket size (1pixel), the algorithm applies a preemptive filtering technique. First, it preempts by identifying the first data point within the bucket. This allows the surrounding data to be filtered based on that data point. The filtering process is dynamic and can adjust based on real-time data variations and resolution changes. The filtered data is then rendered for output. The R-PPF algorithm's strength lies in its ability to adapt to varying data resolutions and dynamically filter out irrelevant or redundant pixels, resulting in a cleaner dataset and much faster visualization for further analysis or application.

This method retrieves the entire data from the database and processes it on the client side; suffer from lengthy data retrieval times. To overcome this, we introduce the Advanced

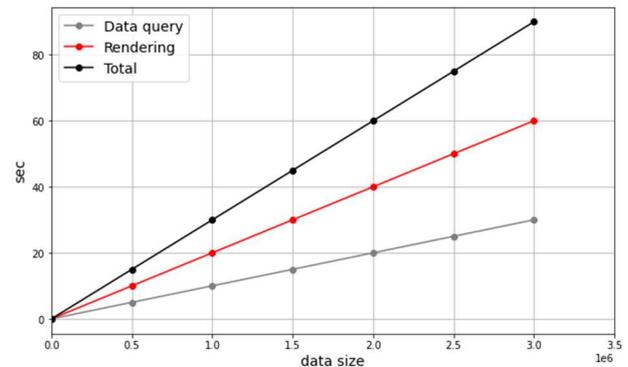

Fig. 1: Time-Series Data Retrieval and Charting Time Comparison of InfraEES.

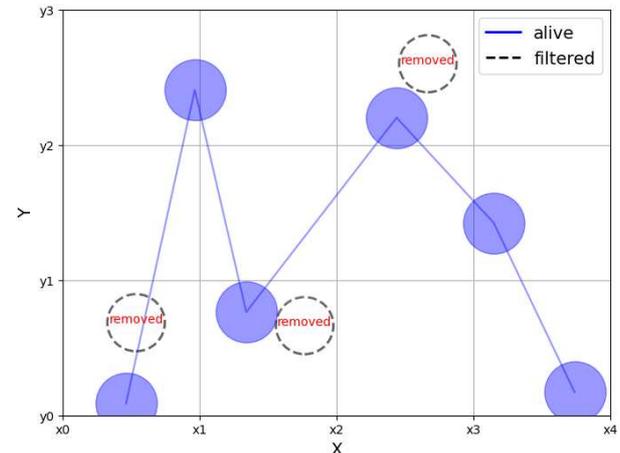

Fig. 2: Data Filtering Process Using the R-PPF Algorithm.

Resolution-based Pixel Preemption Data Filtering (AR-PPF) algorithm. This algorithm initiates a pre-processing stage within the database, where it first filters and stores the data according to predefined criteria. Subsequently, during the client-side processing, it further refines the data based on the user-defined chart query range. This two-tiered approach ensures that the data retrieved and processed on the client side is already optimized for efficient rendering, thereby improving performance and reducing the overall processing time. Consequently, the time required for data visualization over a one-month query period is significantly reduced from 20 seconds to just 0.8 seconds.

[1]MES Team, Innovation Center, SEC

## 2 Related Work

In terms of a time series-based chart, the process of preemptively eliminating data points from a visual perspective can be construed as a method for line simplification [3]. It involves reducing the number of points in a geometric shape while maintaining its original form as much as possible.

The Visvalingam-Whyatt algorithm [4] is a technique used for this purpose, particularly in cartographic generalization. It works by assigning importance levels to points based on local conditions, then removing points in order of least to most importance. The importance is determined by the triangular area each point contributes. The algorithm continues to remove points until all generated triangles are larger than a predetermined threshold. However, it may not distinguish between sharp spikes and shallow features, potentially eliminating important details. Additionally, it is less efficient in terms of processing speed compared to other similar algorithms.

The Douglas-Peucker algorithm, another line simplification technique referenced in [5], effectively omits superfluous points in line charts. The algorithm begins with a curve, which is an ordered set of points or lines with a given distance dimension $\varepsilon > 0$. The algorithm proceeds by recursively dividing the line. It starts by marking the first and last points of the curve to keep. Then, it identifies the point that is farthest from the line segment with the first and last points as end points. If this point is closer than $\varepsilon$ to the line segment, all unmarked points can be discarded without making the simplified curve worse than $\varepsilon$. This process continues until all points are either kept or discarded, resulting in a simplified version of the original curve.

There are a number of distance-based methods to measure the difference between the simplified data set and the original data [6]. In this paper, we intend to measure the distance by applying the Hausdorff distance. When trying to measure the distance between a set of two points, it cannot be assumed that the points from each set are close to all other points in the other set just because the distance between any two arbitrary points is small. Therefore, instead of using the minimum distance of two points, the maximum distance is used to calculate the distance between the two sets. The Hausdorff distance is a method for similarity measurement employing this concept [7].

The Hausdorff distance between two finite point sets $X$ and $Y$ is defined as follows:

$$H(X,Y) = \max(h(X,Y), h(Y,X))$$

Where $h(Y,X)$ denotes the greatest of all the distances from a point in $X$ to the closest point in $Y$. Here, $x$ represents an element of $X$, $y$ is an element of $Y$, and $\text{dist}(x,y)$ is the distance between $x$ and $y$. Notably, when $X$ is a set of original data points and $Y$ is a simplified subset of $X$, $h(Y,X)$ is always 0. As the Hausdorff distance is always equal to or greater than 0, it can be expressed mathematically as follows:

$$H(X,Y) = h(X,Y) = \max_{x \in X}(\min_{y \in Y}(\text{dist}(x,y)))$$

When comparing the similarity of line charts with emphasized data points, the Hausdorff distance should be calculated considering only the raw data points, excluding the points included in the line. Essentially, this is akin to computing the Hausdorff distance for a scatter plot.

According to this approach, the distance is determined by finding the nearest point in the simplified subset $Y$ for each point $x$ in the original set $X$, and then identifying the point $x$ with the greatest such distance. This is similar to identifying the point x that is not included in the filtered subset $Y$, which becomes apparent when the smallest circle that includes at least one point from set $Y$ is drawn with $x$ as the center. The radius of the largest such circle is the Hausdorff distance. In other words, when compared to the original set $X$, the area around the point $x$ where the largest circle is drawn appears visually empty as there are no remaining points from set $Y$. This method effectively measures the maximum deviation introduced by the simplification process.

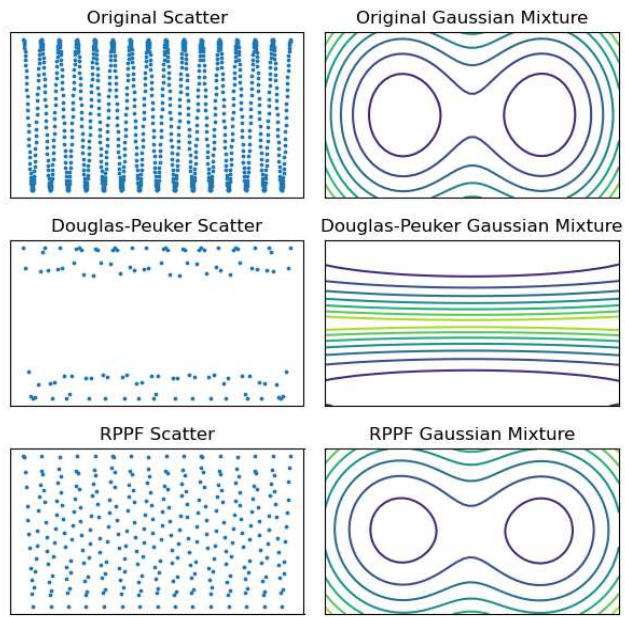

Fig. 3: Comparative Visualization of Original Data, Douglas-Peucker Algorithm, and R-PPF Algorithm through Line and Scatter Plots.

The Douglas-Peucker algorithm eliminates all points within a distance dimension $\varepsilon$ from the line segment, leaving only the endpoints. The point placed at the center of the largest empty circle among the removed points indicates that the Hausdorff distance is half of the original length of the line segment. This represents the most extreme case of dissimilarity between the subsets before and after the application of the Douglas-Peucker algorithm.

Using the R-PPF algorithm, only one point among the points classified into the same bucket is left, so the Hausdorff distance is equal to or less than the maximum length of the bucket space. Therefore, if the bucket size is set to be sufficiently small in a scatter chart, a lower Hausdorff distance can be obtained compared to the Douglas-Peucker algorithm.

Fig. 3 provides a side-by-side comparison of scatter and Gaussian mixture distribution plots for the Original Data, Douglas-Peucker algorithm, and R-PPF algorithm. The representation of each dataset is demonstrated in separate rows, with corresponding scatter plots and distribution plots. The plots generated using the Douglas-Peucker algorithm and the R-PPF algorithm demonstrate their line simplification capabilities by strategically omitting specific data points from the original dataset.

The Douglas-Peuker algorithm omits the majority of data points in linear regions. This results in a substantial increase in the Hausdorff distance in these areas when compared to the original data, suggesting a significant data reduction in these regions. In addition, as can be seen in the second column, it is evident that the distribution itself is distorted. In contrast, the RPPF algorithm uses a bucket-based approach for data sampling. As a result, the gaps between the data points displayed in the plot do not exceed twice the size of the bucket, providing a more balanced data representation.

## 3 Method

The previously introduced R-PPF still faces some challenges. R-PPF performs the task of simplifying data in the user interface (UI) after querying the entire original data from the database (DB) as the user has queried. In other words, if the query period is long, the time taken increases proportionally as the size of original data increases. If the query period exceeds a certain range, users may still feel it is slow. To improve this, we propose a method to pre-determine the bucket size, store the simplified data into the database in advance, and then query this preprocessed data in the UI. This approach significantly reduces the data volume compared to querying the original dataset.

To determine an optimal bucket size that precludes any noticeable visual difference to users, it is essential to have knowledge of the queried data's period and the maximum and minimum values within that period. However, these factors are only confirmed after the chart has been fully rendered. If the bucket size is pre-set for batch processing, the resulting chart shape may deviate from the one created on demand.

To compare the Hausdorff distance between the results of the two methods, we will denote the size of the time axis of the bucket as $T$ and the size of the value axis as $V$. The R-PPF algorithm retains only one point within the same bucket. Consequently, all points x that are removed within a bucket are closest to the point that has claimed the bucket, without considering any other points outside the bucket. In other words, if we denote the original point within the same bucket as $x_b$ and the preempted point as $y_b$, The upper bound of the distance from $x_b$ to its nearest point is the distance to $y_b$. Particularly, as $x_b$ and $y_b$ are within the same bucket which is in the shape of a rectangle, the points are farthest apart when they are at the opposite ends of the diagonal as $D$. This can be summarized as follows:

$$\min_{y \in Y}(\text{dist}(x_b, y)) \leq \text{dist}(x_b, y_b) \leq \sqrt{T^2 + V^2} = D$$

Utilizing this, we can calculate the upper bound of the Hausdorff distance after the R-PPF algorithm is applied. The upper bound of the distance to the nearest $y$ point for all $x$ points is $D$. Since the size of all buckets is the same, we can determine the upper bound as follows, which is equal to $D$.

$$H(X, Y) = \max_{x \in X}(\min_{y \in Y}(\text{dist}(x, y))) \leq \max_{x \in X}(D) = D$$

This indicates that the Hausdorff distance is influenced solely by the bucket size, regardless of when the R-PPF is performed. Assume that we know the distances between three different sets $A$, $B$, and $C$, specifically between $A$ and $B$, and between $B$ and $C$. Then, for points $a$, $b$, c existing on a two-dimensional plane, the following statement always holds true according to the Triangle Inequality.

$$\text{dist}(a, c) \leq \text{dist}(a, b) + \text{dist}(b, c)$$

Consequently, the Hausdorff distance between $A$ and $C$ can be simply expressed as follows:

$$H(A, C) \leq \max_{a \in A}\left(\min_{b \in B, c \in C}(\text{dist}(a, b) + \text{dist}(b, c))\right)$$

If $a$ and $b$ are fixed, the $c$ that minimizes $\text{dist}(b, c)$ becomes the $c$ closest to $b$. The upper bound of $\text{dist}(b, c)$ is $H(B, C)$. In summary, it can be stated as follows:

$$H(A, C) \leq H(A, B) + H(B, C)$$

This shows that the Hausdorff distance resulting from executing the R-PPF algorithm on the original dataset $A$, with a pre-determined bucket size to produce Dataset $B$, and then executing the same algorithm with an visually optimized bucket size for the user's queried range, is at most the sum of the distances from $A$ to $B$ and from $A$ to $C$, respectively. Furthermore, since the distance is upper-bounded by the diagonal length $D$ of the bucket size, the distance between $B$ and $C$ cannot exceed the sum of their respective $D$s. This can be expressed as follows:

$$H(B, C) \leq H(A, B) + H(A, C) \leq D_B + D_C$$

The same holds true when the R-PPF algorithm is applied in a chain. The upper bound of the distance before and after the algorithm is applied at each stage is the $D$ used at that time, and the upper bound of the distance between the original dataset and the final result is the total sum of $D$ at all stages. If we denote the $i$-th application result of R-PPF as $B_i$, the total number of applications as $n$, and the diagonal length of the bucket applied at the $i$-th step as $D_i$, it can be expressed as follows:

$$H(A, B_n) \leq H(A, B_1) + \sum_{i=2}^{n} H(B_{i-1}, B_i) \leq \sum_{i=1}^{n} D_i$$

AR-PPF operates under the premise that it can limit the final Hausdorff distance. It preprocesses with R-PPF, stores it in the DB, fetches the simplified data from the UI, and processes it again with R-PPF to ultimately show the user a visually similar chart. There are two important elements for limiting the distance in preprocessing. One is to process the R-PPF in time-axis bucket units to minimize the distance of preprocessing, and the other is multi-pass processing for memory efficiency in real-time processing.

The first element involves appropriately adjusting the batch unit along the time axis when preprocessing with the R-PPF algorithm. Fundamentally, R-PPF determines the size of the visually optimized bucket based on the data existing in the range the user wants to query. This means that the target bucket size is dynamic, and we cannot know how long the range is along the time axis or what the range is along the value axis. If the size of the bucket $D_{\text{pre}}$ used in preprocessing is much larger than the target size of the bucket $D_{\text{target}}$, the user may perceive it as a visually different chart and feel that important information has been omitted.

In order to make $D_{\text{pre}}$ smaller than $D_{\text{target}}$ by dividing the pre-processing bucket size into time axis length $T$ and value axis length $V$, one of $T_{\text{pre}}$ and $V_{\text{pre}}$ must be smaller than $T_{\text{target}}$ and $V_{\text{target}}$, respectively. While it is challenging to limit the lower bound of $V_{\text{target}}$, since it can be 0 in cases where the same value is maintained, the lower bound of $T_{\text{target}}$ can be limited by forcing the original dataset of that range to be queried in the same way as the conventional R-PPF method if the user's query period is within a certain period. We have determined this period to be seven days, which is the limit where the result of the conventional method can be output to the user within one second.

Now, if $V_{\text{pre}}$ can be made smaller than $V_{\text{target}}$, pre-processing can be done while maintaining $D_{\text{pre}}$ smaller than $D_{\text{target}}$. In order to visualize the entire dataset on the chart, the range of the Y-axis is influenced in proportion to the maximum and minimum range of values. This is also the same for $V$. In other words, if we change the above condition differently, the maximum and minimum range used in preprocessing should be smaller than the maximum and minimum range of the range dynamically queried by the user. If the elements of a set $A$ belong to another set $B$, then $B$ also includes the maximum and minimum values in $A$, so $B$ has the same or larger maximum and minimum range than $A$. By leveraging this, if the entire data used in the preprocessing batch is always included in the user's query range, which can change dynamically, $V_{\text{pre}}$ can be limited to be the same or smaller than $V_{\text{target}}$. Therefore, we set one $T_{\text{pre}}$ interval as the batch processing unit and set the maximum and minimum range of data within that interval as $V_{\text{pre}}$. Users are always restricted to query in units of $T_{\text{pre}}$, ensuring that the entire batch interval data is always included in the user's query range, thus always satisfying $D_{\text{pre}} \leq D_{\text{target}}$.

To be precise, for the $n$ batch-processed datasets included in the final result $B_{\text{target}}$ that the user sees, if we denote the diagonal length of the bucket used in the $i$-th dataset $B_i$ as $D_i$, the upper limit of $H(A, B_{\text{target}})$ can be expressed as follows:

$$H(A, B_{\text{target}}) \leq \max_{a \in A}(\bigcup_{i=1}^{n} \min_{b \in B_i}(\text{dist}(a, b))) \leq \max_{1 \leq i \leq n} D_i$$

The next important element is to process the same range multiple times in real time using less memory. If the batch is processed in one pass, all sensor data during the $T_{\text{pre}}$ interval must be collected to calculate $V_{\text{pre}}$, occupying memory. The occupied memory size increases as $V_{\text{pre}}$ is longer or the sensor sampling period is shorter. If, after collecting data in one pass, RPPF is applied immediately, stored in memory, and RPPF is applied again to the data collected in the next pass and the result stored in the existing memory, the occupied memory becomes the sum of the number of data collected during the pass and the number of buckets on the value axis. That is, the size of the occupied memory can be limited by adjusting the pass period and the number of buckets.

However, the upper bound of the Hausdorff distance increases gradually as the R-PPF algorithm is applied repeatedly. Assuming that the upper bound is $D$ when processing the batch in one pass, the multiple pass processing adds the previous $D$ to the upper bound $D$ of the last pass. Each pass stores the maximum and minimum values from the previous pass, so it can be said that the previous $D$ is smaller than the next iteration's $D$ as the passes progress. However, in the worst case scenario, final $D$ can increase exponentially with the number of passes. Therefore, the number of passes should be appropriately adjusted to manage the final $D$.

For example, where the pre-processing is limited to every 5 minutes due to memory size limitations, the same bucket can be repeatedly processed using the R-PPF algorithm for each 5-minute interval. The required memory size will be the sum of the number of data collected in 5 minutes and the number of vertical buckets. If $T_{\text{pre}}$ is 30 minutes and each pass occurs every 5 minutes, it takes a total of 6 passes to complete one batch processing. In this case, if the size of the bucket used is $D$, the worst-case scenario for the final Hausdorff distance would be $D$ multiplied by six.

If the sampling period of the sensor is 1 second, there will be a total of 1,800 data points over a period of 30 minutes. When processed in passes of 5-minute intervals, each pass will consist of 300 new sensor data points and value-axis buckets, even if set to 200, will occupy a total of 500 memory slots. This means that compared to processing the entire batch in one pass, it will only occupy 28% of the memory.

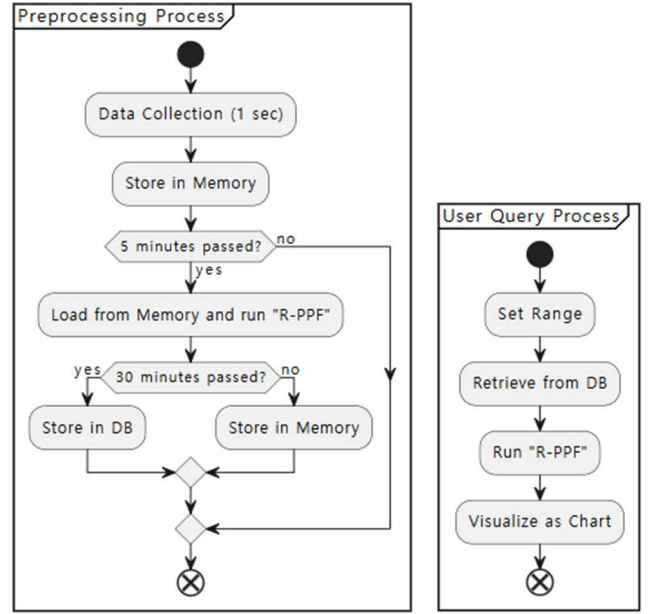

Fig. 4: Flow Chart of AR-PPF process

After handling the two elements mentioned above, the UI retrieves simplified data through pre-processing for the specific interval that the user requested. Finally, this data is further reduced by performing the R-PPF algorithm based on the user's query period (Fig. 4).

The purpose of AR-PPF is to reduce the amount of data queried from the database and used for chart drawing. This approach allows us to efficiently utilize the available memory resources while ensuring an acceptable Hausdorff distance.

## 4 Experiments

### 4.1 Dataset

For ease of calculation, we assume that when the user divides the bucket into a 300 by 100 matrix in the chart they are viewing, points within the same bucket are visually perceived as identical points. We assume that the unit for both the width and height of the bucket is 1, and we omit the units for time or value for convenience. This forces the upper bound of the Hausdorff Distance to be within the diagonal length of the bucket, which is $\sqrt{2}$ or approximately 1.42.

In our experiment, we use 100,000 data points across various dataset types: linear time axis, 0.1 periodic function, 10 periodic function, uniformly distributed sampling, and normally distributed sampling.

Each dataset is presented on a 300 by 100 matrix display with a unit scale of 1, without the application of any simplification algorithms for easy comparison. Hence, the range of all datasets is set between 0 and 100. In the case of normally distributed sampling, outliers that exceed the range may be generated. To address this, any data point that falls outside the 3 sigma radius from the mean of 50 is adjusted to the limit of the range, ensuring that 99.7% of the data falls within the specified limits (described in Table 1).

Table 1: Dataset for the experiment.

| Type | Method |
|---|---|
| Linear time axis | $f(x) = \frac{x}{3}$ |
| 0.1-Periodic Function | $f(x) = 50 \cdot \cos\left(\frac{2\pi x}{0.1}\right) + 50$ |
| 10-Periodic Function | $f(x) = 50 \cdot \cos\left(\frac{2\pi x}{10}\right) + 50$ |
| Uniform Distribution | $x \sim \mathcal{U}(0, 100)$ |
| Normal Distribution | $x \sim \mathcal{N}(50, 50/3)$ |

## 4.2 Experiment Setting

We conducted three experiments to compare the performances of several methods. In the first experiment, we compared the Hausdorff Distance values between the original data and the simplified data obtained by applying the Douglas-Peucker algorithm and R-PPF based algorithms to each dataset. The smaller the distance in this experiment, the closer the simplified data is considered to be visually similar to the original data. If the distance is within 1.42, users may perceive no visual difference. We compared the distance by Douglas-Peucker algorithm with epsilon; filtering boundary; set to 1.4, the Douglas-Peucker algorithm with epsilon set to 0.7. According to the definition of the algorithm, a smaller epsilon value should result in a smaller distance, indicating a closer similarity to the original data. Therefore, if the epsilon value is set to 0.7, the distance should be smaller than if it were set to 1.4.

The R-PPF based algorithms inherently perform R-PPF on the same interval in the final step, which means that the simple R-PPF will show the lowest distance. On the other hand, AR-PPF is expected to have an increasing upper bound on the distance as the number of passes increases, resulting in an increase in the distance as well. AR-PPF uses the bucket length of a time axis as the unit for batch processing. A pass refers to the number of times R-PPF is performed in one batch processing during the preprocessing stage. It processes the number of data included in time axis length 1, divided by the number of passes, to ensure the upper limit of memory occupancy is uniform for each pass. In a 1-pass process, all data within a time unit occupy the memory. However, a 5-pass process divides this data into 5 equal sets. Each pass then uses memory for both the divided data and the results from the previous pass.

The next experiment compares the total number of data points that need to be queried from the database, the Reprocess Batch Size, for the R-PPF based algorithm. The R-PPF algorithm queries the entire dataset, while the AR-PPF based algorithm retrieves simplified data through pre-processing, resulting in fewer data points being queried. With fewer data points, the database query time is reduced, leading to improved query speed.

We compare the occupied memory size, the Pre-process Batch Size, based on the number of passes in AR-PPF. In this experiment, as the number of passes increases, the occupied memory size should be smaller. As a result, a smaller occupied memory size allows to process more sensors within the same memory resources.

## 4.3 Result

As shown in Fig. 5, the use of the Douglas-Peucker algorithm results in a much larger Hausdorff distance compared to methods based on R-PPF. This is attributed to the omission of all points included in the linear sections.

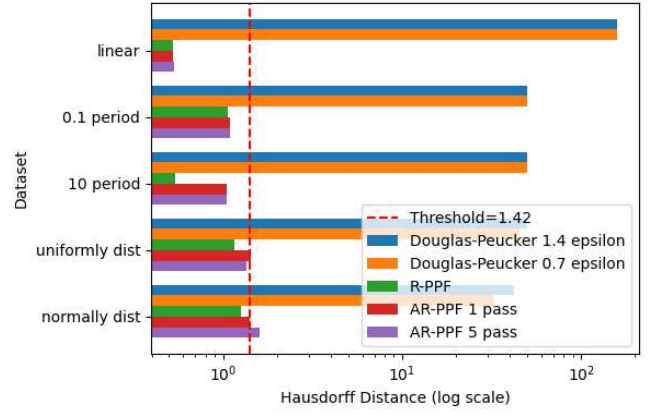

Fig. 6: Hausdorff distance by algorithm and dataset.

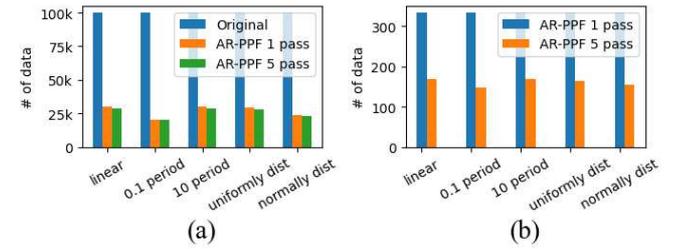

Fig. 5: Performance Comparison Results of the AR-PPF Algorithm. (a) represents the number of data retrieved from DB; (b) represents the number of data occupying memory in each pass.

The threshold, indicated by the red dashed line, represents the standard value of 1.42, which is indistinguishable to the user. Notably, the AR-PPF algorithm with five passes exceeds this threshold in the normally distributed sampling dataset. As explained in the Section 3, when performing five passes, the upper limit of the distance is the sum of the diagonal lengths of the buckets used in each pass. The distance at this point is 1.602, and since all operations used buckets of the same size, it can be observed that this is significantly smaller compared to the upper limit.

When comparing the distance results between algorithms based on R-PPF, AR-PPF always yields a larger value than R-PPF. The difference between these two lies in the preprocessing of data; at the point when a user retrieves data from the UI, the same R-PPF algorithm is applied to the data retrieved from the DB. Therefore, the upper bound of the distance inherently includes that of the R-PPF algorithm, and AR-PPF has an increased upper bound due to preprocessing. In a uniformly distributed sampling dataset, the distance result of a 1-pass process in AR-PPF is larger than that of a 5-pass process. This demonstrates that we can only limit the upper bound of the distance by the number of passes, and we cannot control the actual distance value.

Fig. 6 represents performance comparison results. The chart on the left (a) shows the number of data points queried by each algorithm when a user retrieves data from the UI. Both the Douglas-Peucker algorithm and the R-PPF algorithm query all the data since they process the data after it is retrieved from the DB. In other words, they both query the same amount of data as the original data. On the other hand, AR-PPF loads reduced data into the DB due to preprocessing and queries this data again, resulting in much fewer number. From the original 100k data points, all datasets require only 25k points, or 25% of the data. This can lead to a DB query speed that is approximately 75% faster than the original, proportional to the number of data points queried.

The AR-PPF values in the chart (a) represent the amount of data stored in the DB, thus it can be seen as the additional DB storage size required to operate the AR-PPF algorithm. Since AR-PPF leaves only one piece of data in each bucket after batch processing, up to 100 value axis buckets are stored for time axis length 1 when the number of data points from the results is the highest. Therefore, the upper limit of the data proportion to be stored is 100 out of 334 data points in time axis length 1 based on the experimental dataset. Hence, the upper limit of the additional DB storage space required is 30% of the original data capacity.

The chart on the right (b) represents the maximum number of data points occupying memory in each pass during batch processing. The total number of data used in a single batch process is 334, which is the same as the memory occupancy in a 1-pass process. In a 5-pass process, this is divided by 5, resulting in the use of 67 data points. Additionally, the number of result data points after each pass process must be stored in memory. In cases where the number of result data points is the highest, it is equivalent to the number of value axis buckets, so when combined, the upper limit of memory occupancy per pass is 167. As can be seen in the chart, there are many cases where the occupancy is lower than the upper limit, which varies depending on the characteristics of the original data. Based on the calculated upper limit and experimental results, increasing from 1-pass to 5-pass can optimize memory occupancy by about 50%. If the number of data used in batch processing is much larger, the influence of the bucket count decreases, making it much more effective to reduce.

## 5 Discussion

In the Section 4.3, we confirmed that by setting an appropriate bucket size through AR-PPF algorithm, charts can be queried at high speeds without visual differences. Additionally, we observed that in server environments with limited memory usage, setting an appropriate bucket size and number of passes can reduce memory consumption while limiting the upper bound of any additional visual discrepancies.

In Fig. 7, the chart on (a) displays the data from a specific sensor for one month on the UI screen. It took 202.8 seconds to generate this output. The displayed data includes 30 days of data at one-second intervals, with a total of 390,000 data points queried from the database, excluding the time periods with no data.

When on-demand R-PPF is performed after the query, the speed increases by 90%, taking 20.3 seconds. The chart on (b) shows the result of the AR-PPF algorithm performed at 5-minute intervals, or 6 passes, which took 0.83 seconds to generate. The number of data points queried from the database was only 25% of the original amount. You would not perceive any significant visual differences between charts (a) and (b).

In a project predicting associated sensors based on the 1-second trace chart view of the fault detection and classification (FDC) system, there was a problem of excessively large training data size when using the original data. This issue was addressed by simplifying the training data through the execution of the AR-PPF algorithm, specifying the period that users frequently query. After training with a clustering algorithm and comparing with actual associations when randomly selected, an accuracy of 91.3% was achieved. Thus, the AR-PPF algorithm can also be employed in machine learning training, where it has proven effective in reducing the training time efficiently.

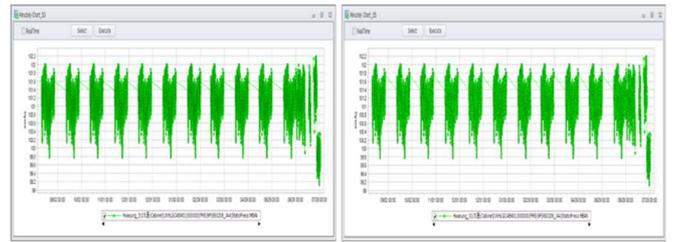

(a)          (b)

Fig. 7: Comparison of the chart screens between the original data and the AR-PPF algorithm in the UI.

The AR-PPF algorithm has been implemented across multiple systems. In the EES UI system, it underwent a trial using data collected from 0.1-second sensor readings over approximately 1.5 days. The results demonstrated a 93% improvement in query speed, reducing the original 114 seconds to 8.78 seconds. This validates the versatile applicability of the AR-PPF algorithm and confirms the performance enhancements it brings.

## 6 Conclusion

As smart factories become more prevalent, there is an increasing demand for visualizations that enable long-term time-series data analysis. Our proposed AR-PPF algorithm addresses this need by allowing users to retrieve and visualize data for a month within one second, even with limited memory. Furthermore, the algorithm produces results that are visually very similar to the original data, with Hausdorff distance that can be adjusted according to the bucket size.

While our approach offers significant improvements over existing methods, there are still a room for improvement. Our algorithm has the same limitation as before when the entire period exceeds a certain interval compared to the bucket size. In order to overcome this limitation, we plan to explore ways to hierarchically preprocess each unit interval of the resolution, regardless of the size of the entire period. Then the entire data can be visualized in a single chart.